\documentclass[english,preprint,tightenlines,showpacs,showkeys,floatfix, 
nofootinbib,superscriptaddress,fleqn]{revtex4}
\usepackage[T1]{fontenc}
\usepackage[latin9]{inputenc}
\usepackage{float}
\usepackage{bm}
\usepackage{graphicx}

\makeatletter

\providecommand{\tabularnewline}{\\}

\makeatletter
\usepackage{graphics}
\usepackage{epsf}
\usepackage{bm}
\usepackage{color}

\makeatother

\usepackage{babel}

\begin{document}
\preprint{INHA-NTG-05/2010}

\title{Tensor charges and form factors of SU(3) baryons in the self-consistent
  SU(3) chiral quark-soliton model}

\author{Tim Ledwig}

\email{ledwig@kph.uni-mainz.de}

\affiliation{Institut für Kernphysik, Universität Mainz, D
-55099 Mainz, Germany} 

\author{Antonio Silva}

\email{ajsilva@fe.up.pt}

\affiliation{Faculdade de Engenharia da Universidade do Porto,
  R. Dr. Roberto Frias s/n, P-4200-465 Porto, Portugal} 

\affiliation{Centro de Fisica Computacional (CFC), P-3004-516 Coimbra, 
  Portugal}

\author{Hyun-Chul Kim}

\email{hchkim@inha.ac.kr}

\affiliation{Department of Physics, Inha University, 
Incheon 402-751, Korea}

\date{June, 2010}

%
%
%
%

\begin{abstract}
We investigate the tensor form factors of the baryon octet within the
framework of the chiral quark-soliton model, emphasizing those of the
nucleon, taking linear $1/N_c$ rotational as well as linear $m_s$
corrections into account, and applying the symmetry-conserving
quantization. We explicitly calculate the tensor form 
factors $H_{T}^{q}(Q^{2})$ corresponding to the generalized parton
distributions $H_{T}(x,\xi,t)$. The tensor form factors are obtained
for the momentum transfer up to $Q^{2}\leq1\,\mathrm{GeV}^{2}$ and at
a renormalization scale of $0.36\,\mathrm{GeV}^{2}$. We find for the
tensor charges $\delta u=1.08$, $\delta d=-0.32$, $\delta
s=-0.01$ and discuss their physical consequences, comparing them with
those from other models. Results for tensor charges for the baryon 
octet are also given.
\end{abstract}

\pacs{13.88.+e, 12.39.-x, 14.20.Dh, 14.20.Jn}

\keywords{Form factors, chiral quark-soliton model, tensor charges,
transversity}

\maketitle

%
%
%
%

\section{Introduction}
At leading twist, the basic quark structure of the nucleon is 
described by three independent quark parton distribution functions
(PDFs): The unpolarized PDF $q(x)$, the helicity PDF $\Delta q(x)$,
and transversity PDF $\delta q(x)$ for each flavor. More generally,
eight generalized parton distributions (GPDs) contain full information
on the structure of the nucleon: Two chiral-even spin-independent GPDs 
$H(x,\xi,t)$ and $E(x,\xi,t)$,  two chiral-even spin-dependent GPDs
$\tilde{H}(x,\xi,t)$ and $\tilde{E}(x,\xi,t)$, and four chiral-odd
spin-dependent GPDs $H_T(x,\xi,t)$, $E_T(x,\xi,t)$
,$\tilde{H}_T(x,\xi,t)$, and
$\tilde{E}_T(x,\xi,t)$\cite{Ralston:1980pp,Mueller:1998fv,Ji:1997,
  Hoodbhoy:1998,Diehl:GPDhFlip}.  Thus, knowing all these
leading-twist GPDs will provide a detailed 
description of the nucleon structure, so called a \textit{nucleon 
 tomography}.

The helicity distributions are related to the axial-vector current of
the nucleon, which are relatively easily accessible in experiment because
of their chiral-even character. On the contrary, the chiral-odd 
distributions that are pertinent to the tensor current of the nucleon
are rather difficult to be measured. Since quantum chromodynamics
(QCD) possesses an approximate chiral symmetry and all electroweak
vertices preserve chirality, properties of the tensor current are
experimentally very hard to be accessed. There exist no probes to
measure directly the tensor structure, so that one is restricted to
scattering reactions in which two chiral-odd processes are
involved. For example, the transverse spin asymmetry $A_{TT}$ in
Drell-Yan processes in $p\overline{p}$ reactions 
\cite{Efremov:2004,Anselimo:2004,PAX:2005,Pasquini:2006} as well as  
the azimuthal single spin asymmetry in semi-inclusive deep inelastic
scattering (SIDIS) \cite{Anselmino:tensorcharge} are seen as promising 
reactions to gain information on the transversity of the nucleon.

Because of this difficulty, an experimental extraction of the
transversity distribution $\delta q(x)$ was only recently performed
for the first time ~\cite{Anselmino:tensorcharge}. Based on 
the azimuthal single spin asymmetry in SIDIS processes
$lp^{\uparrow}\to l\pi X$, Ref.~\cite{Anselmino:tensorcharge} used
experimental data from the Belle Collaboration at KEK
\cite{Belle:2006} as well as data from the HERMES~
\cite{HERMES:2005,HERMES:2006} and COMPASS~\cite{COMPASS:2007}    
collaborations in order to obtain results for
the $u$ and $d$ quark transversity distributions $\delta u(x)$ and
$\delta d(x)$. Subsequently, Ref.~\cite{Cloet:tensorcharge}
extracted the tensor charges $\delta u=0.46_{-0.28}^{+0.36}$ and
$\delta d=-0.19_{-0.23}^{+0.30}$ at a renormalization scale of
$\mu^{2}=0.4\,\mathrm{GeV}^{2}$. Anselmino et al.~\cite{Anselmino:2009}
presented an updated analysis and obtained the results $\delta 
u=0.54_{-0.22}^{+0.09}$ and $\delta d=-0.23_{-0.16}^{+0.09}$ 
at a scale of $\mu^{2}=0.8\,\mathrm{GeV}^{2}$.  Moreover, several
theoretical investigations on the tensor charges were carried out, for  
example, in the non-relativistic quark model, in the MIT bag
model~\cite{HeJi:tensor}, in SU(6)-symmetric quark 
models~\cite{Schimdt:HOmodel,Pasquini:2005}, in a valence quark model
with axial-vector meson dominance~\cite{Gamberg:2001qc}, and on the  
lattice~\cite{Goeckeler:GPDlattice}.   

In the present work, we will study the tensor form factors
up to a momentum transfer of $Q^{2}\leq 1\,\mathrm{GeV}^{2}$ within 
the framework of the self-consistent SU(3) chiral quark-soliton model 
($\chi$QSM)~\cite{Christov:1995vm}. The $\chi$QSM provides an
effective chiral model for QCD in the low-energy regime with
constituent quarks and the pseudo-scalar mesons as the 
relevant degrees of freedom. Moreover, this model has a deep
connection to the QCD vacuum based on
instantons~\cite{CQSM:Diakonovlecture} and has only a few free 
parameters. These parameters can mostly be fixed to the meson 
masses and meson decay constants in the mesonic sector. The only free 
parameter is the constituent quark mass that is also fixed by
reproducing the proton electric form factor.  We obtain numerically an
explicit (self-consistent) pion-field, i.e. the non-topological
soliton field, which can be used to calculate basically all baryon
observables. A merit of this model is that we are able to  
determine baryon form factors of various currents within exactly the
same relativistic framework and with the same set of
parameters. Furthermore, the renormalization scale for the $\chi$QSM
is naturally given by the cut-off parameter for the regularization
which is about $0.36\,\mathrm{GeV^2}$. Note that it is implicitly
related to the inverse of the size of instantons
($\overline{\rho}\approx 0.35\,\mathrm{fm}$)
~\cite{Diakonov:1983hh,Diakonov:1985eg}.  In particular, this
renormalization scale is of great importance in investigating the
tensor charges of the proton. In contrast to the axial-vector charges
the tensor charges depend on the renormalization scale already at
one-loop level. 

The axial-vector form factors were calculated in
Refs.~\cite{Silva:AxialFF,Silva:G0A4Happex,Silva:ParityViol,
  Silva:StrangeFF} with the same parameters 
as used in the present work. The tensor charges, i.e. the tensor form
factors at $Q^2=0$, were already investigated in the SU(2) version of
the $\chi$QSM in Refs.~\cite{TensorSU2,Wakamatsu:H(x00)} and in the
SU(3) version in Ref.~\cite{TensorSU3}. However, the SU(3) 
calculation~\cite{TensorSU3} was done prior to the finding of the   
symmetry-conserving quantization~\cite{SymmetryConQuantization} 
which ensures the correct electromagnetic gauge invariance.
Reference~\cite{CQSM:lightCone} formulated the $\chi$QSM on the 
light-cone, within which the tensor charges were
studied in~\cite{Lorce:tensor}. In the case of the SU(2) $\chi$QSM the
tensor charges were also calculated from the first moments of the
transversity distribution $\delta q(x)$ in the
references~\cite{Gamberg1998:H(x00),Wakamatsu1999:H(x00),Schweitzer2001:H(x00)}.  
In the present work, we will extend
the previous work~\cite{TensorSU3} and calculate the tensor form 
factors up to $Q^2=1\,\mathrm{GeV}^2$ with application of the 
symmetry-conserving quantization.  

The outline of this work is sketched as follows. In Section II we
recapitulate the form factors of the tensor current such that it can
be used in the $\chi$QSM. In Section III, we show how to compute the
tensor form factors of SU(3) baryons within the $\chi$QSM. Section IV 
presents the main results of this work and discusses them in
comparison with those of other works. The last Section is devoted to a
summary and conclusion.  The explicit expressions for the quark
densities are given in the Appendix.

%
%
%
%

\section{General Formalism}
The matrix element of the quark tensor current 
between two nucleon states is parametrized by three form
factors~\cite{Diehl:GPDhFlip,Goeckeler:GPDlattice,Hagler:FFdecomposition}
as follows 
\begin{eqnarray}
\langle N_{s^{\prime}}(p^{\prime})| 
\overline{\psi}(0)i\sigma^{\mu\nu} \lambda^\chi\psi(0)
|N_{s}(p)\rangle & = & \overline{u}_{s^{\prime}}(p^{\prime})
\left[H_{T}^\chi(Q^{2}) i   \sigma^{\mu\nu} \;+\;
  E_{T}^\chi(Q^{2})\frac{\gamma^{\mu}q^{\nu} \;-\; 
    q^{\mu}\gamma^{\nu}}{2M}\right. \cr
&& \left. + \;\tilde{H}_{T}^\chi(Q^{2})
  \frac{(n^{\mu}q^{\nu}-q^{\mu}n^{\nu})}{2M^{2}}\right]u_{s}(p)\,, 
\label{eq:general_decomposition}  
\end{eqnarray}
where $\sigma_{\mu\nu}$ is the spin operator 
$i[\gamma_\mu,\,\gamma_\nu]/2$ and $\lambda^\chi$ the Gell-Mann matrices  where we include the unit matrix $\lambda^0=1$. 
 The $\psi$ represents the quark field. The $u_{s}(p)$
denotes the spinor for the nucleon with mass 
$M$, momentum $p$ and the third component of its spin $s$. The
momentum transfer $q$ and the total momentum are defined respectively
as $q=p^{\prime}-p$ with $q^{2}=t=-Q^{2}$ and $n=p^{\prime}+p$. 
In the language of GPDs the above-defined form factors are equivalent 
to the following expressions\footnote{In the notation of the
  generalized form factors of \cite{DiehlHaegler:ATMM} the above form
  factors are equivalent to $H_{T}^{\chi}(q^2)=A_{T10}^\chi(t)$,
  $E_T^\chi(q^2)=B_{T10}^\chi(t)$ and
  $\tilde{H}_T^\chi(q^2)=\tilde{A}_{T10}^\chi(t)$.}:   
\begin{equation} 
\int_{-1}^1dx\, H_T^\chi(x,\xi,t)\;=\; H_T^\chi(q^2),\;
\int_{-1}^1dx\, E_T^\chi(x,\xi,t) \;=\; E_T^\chi(q^2),\;
\int_{-1}^1dx\, \tilde{H}_T^\chi (x,\xi,t)\;=\; \tilde{H}_T^\chi(q^2).
\end{equation}
In the present work, we will concentrate on the tensor
form factors $H_{T}(Q^{2})$ that can be related to the spatial part of
the above matrix element in the Breit frame
\begin{equation}
\label{eq:EnklTkl}
\varepsilon^{nkl}\langle
N_{s^{\prime}}^{\prime}|\overline{\psi}(0)i\sigma_{kl} \lambda^\chi \psi(0)|N_{s}\rangle 
\;=\; H_{T}^\chi(Q^{2})i2\phi_{s^{\prime}}\left[\sigma^{n}+q^{n}
  \frac{{\bm q} \cdot {\bm\sigma}}{4M(E+M)}\right]\phi_{s}  
\end{equation}
with $E=\sqrt{M^{2}+{\bm p}^{2}}$ as the nucleon energy, $\phi_{s}$
the two-component spinor and $N^{\prime}=N(p^{\prime})$. In order 
to derive the expressions for $H_T^\chi(Q^2)$, we take the third
component of the space, i.e. $n=3$ and perform an average over the
orientation of the momentum transfer. Then we act first
$\int\frac{d\Omega}{4\pi}[{\bm q}\times({\bm q}\times$ on both sides,
and take the average. The results are found to be 
\begin{eqnarray}
\int\frac{d\Omega_{q}}{4\pi}\langle
N_{\frac{1}{2}}^{\prime}|{T}_{z}^\chi| N_{\frac{1}{2}}\rangle & = &
H_{T}^\chi (Q^{2})\, i\frac{2M+E}{M}\, 
\frac{2}{3}-E_{T}^\chi(Q^{2})\, i\frac{|{\bm q}|^{2}}{M^{2}}\,\frac{1}{3}\,,
\label{eq:T1}\\ 
\int\frac{d\Omega_{q}}{4\pi}\left[{\bm q}\times \left({\bm q}\times
    \langle N_{\frac{1}{2}}^{\prime}|{\bm
      T}|N_{\frac{1}{2}}\rangle\right)\right]_{z} & = &
-H_{T}^\chi(Q^{2})\, i|{\bm
  q}|^{2}\frac{4}{3}+E_{T}^\chi(Q^{2})\frac{i}{M^{2}}|{\bm
  q}|^{4}\frac{1}{3}
\label{eq:T2}
\end{eqnarray} 
with ${\bm T}^\chi = i\varepsilon^{nkl} \overline{\psi}(0) \sigma_{kl} \lambda^\chi \psi(0) \hat{\bm e}_{n}$. We can therefore
separate $H_T^\chi$ and $E_T^\chi$ as    
\begin{eqnarray}
H_{T}^\chi(Q^{2}) & = & 3\frac{M}{E}\,\,\int\frac{d\Omega_{q}}{4\pi}
\left\{ {\bm T}_{NN}^{\chi} + \frac{1}{|{\bm q}|^{2}} \left[{\bm q}
    \times \left({\bm q} \times{\bm T}_{NN}^\chi\right)
  \right]\right\}_z\,,
\label{eq:HTforModel}\\
E_{T}^\chi (Q^{2}) & = & \frac{12M^{3}}{E|{\bm q}|^{2}} \int
\frac{d\Omega_{q}}{4\pi} \left\{ {\bm T}_{NN}^{\chi} + \frac{2M+E}{2M|
    {\bm q}|^{2}}\left[{\bm q}\times\left({\bm q}\times{\bm T}_{NN}^\chi
    \right )\right]\right\}_z \,,
\label{eq:ETforModel}
\end{eqnarray}
using the relation $i\varepsilon^{nkl}\sigma_{kl} =
2i\gamma^0\gamma^{n}\gamma^{5}$ and the definition
\begin{eqnarray}
{\bm T}_{NN}^\chi & := & \langle
N_{\frac{1}{2}}^{\prime}|\psi^{\dagger}(0){\bm
  \gamma}\gamma^{5} \lambda^\chi \psi(0)|N_{\frac{1}{2}}\rangle\,. 
\label{matelement}
\end{eqnarray} 

Equations~(\ref{eq:HTforModel},\ref{eq:ETforModel}) can be now
evaluated in the $\chi$QSM. We want to note that both form factors
involve the expressions $({\bm T}_{NN}^{\chi})_z$ and $\left[{\bm
    q}\times\left({\bm q}\times{\bm  T}_{NN}^\chi\right)\right]_{z}$. 
However, we will concentrate in the
present work on the tensor form factor $H_T^\chi(Q^2)$. Even though both
form factors consist of the same two densities, the second form factor
$E_T(Q^2)$ requires more technical efforts for the region of small 
$Q^2$. In addition, the third form factor, $\tilde{H}_T^\chi(Q^2)$
requires a completely new density.  These two form factors will be
discussed in a future work. At this point we also see explicitly the
difference of the tensor form factors from the axial-vector ones for
which the nucleon matrix element is given as  
\begin{eqnarray}
{\bm A}_{NN}^\chi & = & \langle N_{\frac{1}{2}}^{\prime}|
\psi^{\dagger}(0)\gamma^0{\bm
  \gamma}\gamma^{5} \lambda^\chi \psi(0)|N_{\frac{1}{2}}\rangle\,.
\label{eq:Aaxial}
\end{eqnarray} 
The ${\bm A}_{NN}$ is distinguished from $\bm T_{NN}$ by only a factor
of $\gamma^0$. It indicates that the tensor current turns out to be
anti-Hermitian whereas the axial-vector one is Hermitian. However, in
the nonrelativistic limit the tensor form factors coincide with the 
axial-vector ones, because $\gamma^0$ is replaced by the unity matrix
in this limit.

In the following we will give additional expressions which will be
used later in the present work. With the above defined current we have
the following relations between the individual flavor decompositions
in SU(3) as: 
\begin{eqnarray}
H_T^0(0) = g_T^0 &=& \delta u +\delta d +\delta s    \\ 
H_T^3(0) = g_T^3 &=& \delta u -\delta d   \\
H_T^8(0) = g_T^8 &=& \frac{1}{\sqrt{3}}(\delta u + \delta d -2 \delta s),
\end{eqnarray}
and in SU(2) with $\tau^0=1$:
\begin{eqnarray}
H_T^0(0) = g_T^0 &=& \delta u +\delta d  \\ 
H_T^3(0) = g_T^3 &=& \delta u -\delta d . 
\end{eqnarray}
We want to note that in the literature several notations for the SU(3)
singlet $g_T^0$ and non-singlet $g_T^8$ quantities exist. These are
due to the fact that either $\lambda^0=\sqrt{2/3}\cdot 1$ are chosen
or the factor $\sqrt{1/3}$ is taken out from $g_T^8$. 

Additionionally, in order to compare the present results of the tensor
charges with those of other works, it is essential to know the
renormalization scale. Different values obtained at different scales
can be connected by following NLO evolution
equation~\cite{Evolution1,Evolution2}:
\begin{eqnarray}
\delta q(\mu^{2}) & = &
\left(\frac{\alpha_{S}(\mu^{2})}{\alpha_{S}(\mu_{i}^{2})}\right)^{4/27}
\left[1-\frac{337}{486\pi}
  \left(\alpha_{S}(\mu_{i}^{2})-\alpha_{S}(\mu^{2})\right)\right]
\delta q(\mu_{i}^{2})\,,
\label{eq:evolve}\\ 
\alpha_{S}^{NLO}(\mu^{2}) & = &
\frac{4\pi}{9\ln(\mu^{2}/\Lambda_{\mathrm{QCD}}^{2})} 
\left[1-\frac{64}{81}\frac{\ln\ln(\mu^{2}/\Lambda_{\mathrm{QCD}}^{2})}{ 
    \ln(\mu^{2}/\Lambda_{\mathrm{QCD}}^{2})}\right]
\end{eqnarray} 
with $\Lambda_{\mathrm{QCD}}=0.248$ GeV, the initial
renormalization scale $\mu_{i}^{2}$ and $N_{c}=N_{f}=3$. 

%
%
%
%

\section{SU(3) Chiral Quark-Soliton Model} 
We will now briefly describe the SU(3) $\chi$QSM. We follow the notation used in
Refs.~\cite{AxialTheta,CQSM:ADEFF,CQSM:SHD,CQSM:Theta_VFF} and for 
details we refer to Refs.~\cite{Christov:1995vm,Christov:eleff,Kim:eleff}.  
The SU(3) $\chi$QSM is characterized by the following low-energy 
effective partition function in Euclidean space 
\begin{equation}
\mathcal{Z}_{\mathrm{\chi QSM}} \;=\;
\int\mathcal{D}\psi\mathcal{D}\psi^{\dagger}\mathcal{D}U \exp
\left[-\int   d^{4}x\Psi^{\dagger}iD(U)\Psi\right]=\int\mathcal{D}U
\exp(-S_{\mathrm{eff}}[U])\,, 
\label{eq:part}  
\end{equation}
where $\psi$ and $U$ represent the quark and pseudo-Goldstone boson
fields, respectively. The $S_{\mathrm{eff}}$ denotes the effective
chiral action  
\begin{equation}
S_{\mathrm{eff}}(U) \;=\; -N_{c}\mathrm{Tr}\ln iD(U)\,,
\label{eq:echl}
\end{equation}
where $\mathrm{Tr}$ designates the functional trace, $N_c$ the number
of colors, and $D(U)$ the Dirac differential operator  
\begin{equation}
D(U) \;=\; 
\gamma_{4}(i\rlap{/}{\partial} - \hat{m} - MU^{\gamma_{5}}) =
-i\partial_{4} + h(U) - \delta m
\label{eq:Dirac}  
\end{equation}
with 
\begin{equation}
\delta m  \;=\; \frac{-\overline{m} + m_{s}}{3}\gamma_{4}\bm{1} +
\frac{\overline{m} - m_{s}}{\sqrt{3}} \gamma_{4} \lambda^{8} =
M_{1}\gamma_{4} \bm{1} + M_{8} \gamma_{4} \lambda^{8}\,.
\label{eq:deltam}
\end{equation}
In the present work we assume isospin symmetry, so that the current 
quark mass matrix is defined as
$\hat{m}=\mathrm{diag}(\overline{m},\,\overline{m},\,
m_{\mathrm{s}})=\overline{m}+\delta m$.  
The SU(3) single-quark Hamiltonian $h(U)$ is given by
\begin{equation}
h(U) \;=\;
i\gamma_{4}\gamma_{i}\partial_{i}-\gamma_{4}MU^{\gamma_{5}} -
\gamma_{4} \overline{m}\, ,
\label{eq:diracham}  
\end{equation}
where $U^{\gamma_5}$ represents the chiral field for which we assume
Witten's embedding of the SU(2) soliton
into SU(3)  
\begin{equation}
U^{\gamma_{5}}(x) \;=\; \left(\begin{array}{lr}
U_{\mathrm{SU(2)}}^{\gamma_{5}}(x) & 0\\
0 & 1
\end{array}\right)
\end{equation}
with the SU(2) pion field $\pi^i (x)$ as 
\begin{equation}
U_{SU(2)}^{\gamma_{5}} \;=\; \exp(i\gamma^{5}\tau^i \pi^i(x))
\;=\; \frac{1+\gamma^{5}}{2}U_{SU(2)} + \frac{1-\gamma^{5}}{2}U_{SU(2)}^{\dagger}\,.
\label{eq:embed}
\end{equation}
The integration over the pion field $U$ in Eq. (\ref{eq:part}) can be
carried out by the saddle-point approximation in the large $N_c$ limit
due to the $N_{c}$ factor in Eq. (\ref{eq:echl}).  The SU(2) pion
field $U$ is expressed as the most symmetric hedgehog form 
\begin{equation}
U_{\mathrm{SU2}} \;=\; \exp[i\gamma_{5}\hat{n}\cdot{\bm
  \tau}P(r)]\,  ,
\label{eq:hedgehog}  
\end{equation}
where $P(r)$ is the radial profile function of the soliton.

The baryon state $|B\rangle$ in Eq. (\ref{matelement}) is defined as
an Ioffe-type current consisting of $N_c$ valence quarks in the
$\chi$QSM: 
\begin{equation}
|B(p)\rangle \;=\; 
\lim_{x_{4}\rightarrow-\infty}\,\frac{1}{\sqrt{\mathcal{Z}}}\,  
e^{ip_{4}x_{4}}\,\int d^{3}{\bm x}\, e^{i\,{\bm p}\cdot{\bm x}}\,
J_{B}^{\dagger}(x)\,|0\rangle
\label{eq:Ioffe}  
\end{equation}
with
\begin{equation}
J_{B}(x) \;=\; \frac{1}{N_{c}!}\,\Gamma_{B}^{b_{1}\cdots
  b_{N_{c}}}\,\varepsilon^{\beta_{1}\cdots\beta_{N_{c}}}\,\psi_{\beta_{1}b_{1}}(x)
\cdots\psi_{\beta_{N_{c}}b_{N_{c}}}(x)\,\,\,,   
\end{equation}
where the matrix $\Gamma_{B}^{b_{1}...b_{N_{c}}}$
carries the hyper-charge $Y$, isospin $I,I_{3}$ and spin $J,J_{3}$
quantum numbers of the baryon and the $b_{i}$ and $\beta_{i}$ denote
the spin-flavor- and color-indices, respectively.
Having minimized the action in Eq. (\ref{eq:echl}), we obtain an
equation of motion which is numerically solved in a self-consistent
manner with respect to the function $P(r)$ in
Eq.~(\ref{eq:hedgehog}). The corresponding unique solution 
$U_c$ is called the classical chiral soliton.

So far, we did not introduce any quantum numbers for the system. This is done
by quantizing the rotational and translational zero-modes of the soliton.
The rotations and translations of the soliton are implemented by 
\begin{equation} 
U({\bm x},t)=A(t)U_{c}({\bm x}-{\bm
    z}(t))A^{\dagger}(t)\,,
\end{equation} 
where $A(t)$ denotes a time-dependent SU(3) matrix and ${\bm z}(t)$
stands for the time-dependent translation of the center of mass of
the soliton in coordinate space. The rotational velocity of the
soliton $\Omega(t)$ is now defined as 
\begin{equation}
\Omega=\frac{1}{i}A^{\dagger}\dot{A} =
\frac{1}{2i}\textrm{Tr}(A^{\dagger}\dot{A}\lambda^{\alpha}) 
\lambda^{\alpha}=\frac{1}{2}\Omega_{\alpha}\lambda^{\alpha}.
\end{equation}  
Thus, treating $\Omega(t)$ and $\delta m$ perturbatively with a
slowly rotating soliton assumed and with $\delta m$ regarded as a
small parameter, we derive the collective Hamiltonian of the
$\chi$QSM~\cite{Blotz:1992pw} expressed as 
\begin{eqnarray}
H_{coll} & = & H_{\mathrm{sym}}+H_{\mathrm{sb}}\,\,\,,\label{eq:Ham}\\ 
H_{\mathrm{sym}} & = & M_{c}+\frac{1}{2I_{1}}\sum_{i=1}^{3}J_{i}J_{i}
+ \frac{1}{2I_{2}}\sum_{a=4}^{7}J_{a}J_{a},\\
H_{\mathrm{sb}} & = & \frac{1}{\overline{m}}M_{1}\Sigma_{SU(2)} +
\alpha D_{88}^{(8)}(A) +\beta
Y+\frac{\gamma}{\sqrt{3}}D_{8i}^{(8)}(A)J_{i}\,.
\end{eqnarray}
Diagonalizing the collective Hamiltonian we derive the octet baryon
states 
\begin{equation}
|N_{8}\rangle \;=\; |8_{1/2},N\rangle + c_{\overline{10}}\sqrt{5}|
\overline{10}_{1/2}, N\rangle  + c_{27}\sqrt{6}| 27_{1/2},
N\rangle, 
\label{eq:Nwave}
\end{equation}
where $c_{\overline{10}}$ and $c_{27}$ are mixing parameters expressed
as  
\begin{eqnarray}
c_{\overline{10}}&=& -\frac{I_{2}}{15}\left(\alpha +
  \frac{1}{2}\gamma\right),\;\;\;\;  
c_{27} \; = \; -\frac{I_{2}}{25}\left(\alpha-\frac{1}{6}\gamma\right),
\end{eqnarray}
and $\alpha$ and $\gamma$ represent the effects of SU(3) symmetry
breaking written as 
\begin{eqnarray}
\alpha &=& \frac{1}{\overline{m}}\frac{1}{\sqrt{3}} M_{8}
\Sigma_{SU(2)} - \frac{N_{c}}{\sqrt{3}}M_{8}
\frac{K_{2}}{I_{2}},\;\;\;\; 
\gamma \;= \; -2\sqrt{3}M_{8}\left(\frac{K_{1}}{I_{1}} -
  \frac{K_{2}}{I_{2}}\right)\,.
\end{eqnarray}
The moments of inertia $I_{1}$,$I_{2}$ and $K_{1}$, $K_{2}$ can be
found, for example, in Ref.~\cite{CQSM:SHD}. In the $\chi$QSM the
constituent quark mass $M$ of Eq. (\ref{eq:diracham}) is in
general momentum-dependent and introduces a natural 
regularization scheme for the divergent quark loops in the model. 
However, it is rather difficult to treat the momentum-dependent
constituent quark mass within the present model. Instead, we will take
it as a free and constant parameter, and introduce a regularization
scheme such as the proper-time regularization. 
It is well known that the value of $M=420\,\textrm{MeV}$ together with the
proper time regularization reproduces
very well experimental form factor data for the SU(3) 
baryons~\cite{Christov:eleff,Christov:1995vm,Kim:eleff,Silva:AxialFF,
Silva:ParityViol}. We want to mention that in the calculation of
nucleon structure function the Pauli-Villars regularization is usually
employed~\cite{Weigel:Regularization}. A detailed formalism for the
zero-mode quantization can be found in
Refs.~\cite{Blotz:1992pw,Christov:1995vm,Kim:eleff}.   

%
%
%
%

\section{Tensor Form factors in the Chiral Quark-Soliton Model}  
In this Section we give the final expressions for the tensor form
factor $H_T^\chi(Q^2)$ of Eq. (\ref{eq:HTforModel}) evaluated in the
$\chi$QSM. In the present framework, linear-order corrections coming
from $\Omega(t)$ and $\delta m$ are taken into account while the
translation of the soliton is treated only to the zeroth order.
Keeping the notations of
Refs.~\cite{AxialTheta,CQSM:SHD,CQSM:Theta_VFF}, we find that   
Eq.~(\ref{eq:HTforModel}) turns out to be 
\begin{equation}
H_{T}^{\chi}(Q^{2}) \;=\; \frac{M}{E}\,\,\int dr\,
r^{2}\,\left[j_{0}(|\bm Q|r) \, \mathcal{H}_{T0}^\chi(r) 
\;+\; \sqrt{2}  j_{2}(|{\bm Q}|r)\,  \mathcal{H}_{T2}^\chi(r)  \right]\,\,,
\label{eq:HTModel}
\end{equation}
where the indices $\chi$ denote the singlet ($\chi=0$) and nonsinglet
($\chi=3,\,8$) parts of the tensor form factors. The $j_0(|\bm Q|r)$
and $j_2(|\bm Q|r)$ stand for the spherical Bessel functions. The nucleon
matrix element $\mathcal{H}_{T0}^{\chi}(r)$ is given explicitly in SU(3)
as follows:
\begin{eqnarray}
\mathcal{H}_{T0}^{\chi=3,8} (r) & = &
-\sqrt{\frac{1}{3}} \langle D_{\chi3}^{(8)}\rangle_N \mathcal{A}_{T0}(r)
\;+\; \frac{1}{3\sqrt{3}}\frac{1}{I_{1}} \langle D_{\chi8}^{(8)}
J_{3}\rangle_N \mathcal{B}_{T0}(r) \cr
&& \;-\; \sqrt{\frac{1}{3}}\frac{1}{I_{2}} \langle d_{ab3}D_{\chi
  a}^{(8)}J_{b} \rangle_N \mathcal{C}_{T0}(r) 
\;- \; \frac{1}{3\sqrt{2}}\frac{1}{I_{1}} \langle D_{\chi3}^{(8)} \rangle_N
\mathcal{D}_{T0}(r) \cr
&& \;-\; \frac{2}{3\sqrt{3}} \frac{K_{1}}{I_{1}}M_{8} \langle D_{83}^{(8)}
 D_{\chi8}^{(8)} \rangle_N 
 \mathcal{B}_{T0}(r) + \frac{2}{\sqrt{3}}\frac{K_{2}}{I_{2}}M_{8} \langle
 d_{ab3}  D_{8a}^{(8)}D_{\chi b}^{(8)} \rangle_N \mathcal{C}_{T0}(r) \cr
 &  & -\sqrt{\frac{1}{3}}\left[2M_{1} \langle D_{\chi3}^{(8)}
   \rangle_N +
   \frac{2}{\sqrt{3}}M_{8} \langle D_{88}^{(8)} D_{\chi3}^{(8)}
   \rangle_N \right] \mathcal{H}_{T0}(r) \cr
 &  & +\frac{2}{3\sqrt{3}}M_{8} \langle D_{83}^{(8)}D_{\chi8}^{(8)}
 \rangle_N  \mathcal{I}_{T0}(r) - \frac{2}{\sqrt{3}}M_{8} \langle
 d_{ab3}D_{\chi a}^{(8)} D_{8b}^{(8)} \rangle_N \mathcal{J}_{T0}(r)\, ,
\label{eq:3,8CQSMdensity}\\
\mathcal{H}_{T0}^0 (r) &=& \frac{1}{3}\frac{1}{I_{1}} \langle
J_{3}\rangle_N \mathcal{B}_{T0}(r) -  \frac{2}{3}\frac{K_{1}}{I_{1}}
M_{8} \langle D_{83}^{(8)}\rangle_N  \mathcal{B}_{T0}(r) +
\frac{2}{3}M_{8} \langle D_{83}^{(8)} \rangle_N  \mathcal{I}_{T0}(r)\,,   
\label{eq:0CQSMdensity} 
\end{eqnarray}
where $\mathcal{A}_{T0},\cdots, \mathcal{J}_{T0}$ are the quark densities  
found in the Appendix. The terms with $M_{1}$ or $M_{8}$ are 
strange-quark mass ($m_s$) corrections arising from the operator. The 
operator $J_3$ is the third component of the spin operator. The
$D^{(8)}$ represent the SU(3) Wigner functions and $\langle \rangle_N$
are their matrix elements sandwiched between the collective nucleon
wave functions given in Eq. (\ref{eq:Nwave}).  The results of these
matrix elements are finally given in terms of the SU(3) Clebsch-Gordan 
coefficients. 

From the above SU(3) expressions,
Eqs. (\ref{eq:3,8CQSMdensity},\ref{eq:0CQSMdensity}), we can deduce
straightforwardly the corresponding expressions for the SU(2)
version. Since there is no strange quark in SU(2), most of the above
terms are not present and
Eqs. (\ref{eq:3,8CQSMdensity},\ref{eq:0CQSMdensity}) are reduced in SU(2)
to the following isovector and isosinglet expressions:
\begin{eqnarray}
\mathcal{H}_{T0}^{3} (r) & = &
-\sqrt{\frac{1}{3}} \langle D_{33}\rangle_N \mathcal{A}_{T0}(r)
\;- \; \frac{1}{3\sqrt{2}}\frac{1}{I_{1}} \langle D_{33} \rangle_N
\mathcal{D}_{T0}(r) \label{eq:3CQSMdensitySU2} \\
\mathcal{H}_{T0}^0(r) &=& \frac{1}{3}\frac{1}{I_{1}} \langle
J_{3}\rangle_N \mathcal{B}_{T0}(r),  
\label{eq:0CQSMdensitySU2} 
\end{eqnarray}
with $\langle D_{33}\rangle_N=-1/3$ and $\langle J_{3}\rangle_N=1/2$ as the
corresponding SU(2) matrix elements.

The matrix element $\mathcal{H}_{T2}^\chi$ can be expressed in the
same form of Eqs. (\ref{eq:3,8CQSMdensity},\ref{eq:0CQSMdensity}) with
the operators in the densities of
Eqs. (\ref{eq:3,8CQSMdensity},\ref{eq:0CQSMdensity}) replaced as
descibed in the Appendix.   

At this point we want to mention that the densities
$\mathcal{A}_T(r),\cdots, \mathcal{J}_T(r)$ are similar to those for
the axial-vector form factors
$\mathcal{A}(r),\cdots,\mathcal{J}(r)$~\cite{Silva:AxialFF}. The only   
difference comes from the $\gamma_0$ ($\gamma^4$ in Euclidean space)
in Eqs. (\ref{matelement},\ref{eq:Aaxial}). This results in the fact
that the reduced matrix elements of the operators occurring in
Eq. (\ref{eq:3,8CQSMdensity}) are the same for 
both tensor and axial-vector densities. The difference in
the complete densities is therefore a minus sign in the lower Lorentz
structure. The factor $\gamma_4$, however, makes the densities for the
tensor charges totally different from the axial-vector ones.  The
effective chiral action expressed in Eq. (\ref{eq:echl}) contains in
principle all order of the effective chiral Lagrangians.  Its
imaginary part generates the well-known Wess-Zumino-Witten 
Lagrangian~\cite{Diakonov:1987ty,Christov:1995vm}. In 
order to produce this Lagrangian correctly, one should not regularize
the imaginary part if the momentum-dependence of the constituent quark
mass is turned off. Thus, the contributions from the imaginary part
of the action to an observable do not have any regularization. As for
the tensor charges, it is the other way around, that is, the real part
contains no regularization but the imaginary part. This is due to the
presence of the factor $\gamma_4$ in the tensor operator that switches
the real and imaginary parts.  

%
%
%

\section{Axial-Vector Form factors in the Chiral Quark-Soliton Model}  
In this Section we will discuss shortly the axial-vector form factors
$G_A^\chi(Q^2)$ calculated in the $\chi$QSM. The general baryon matrix
element is decomposed into its Lorentz-structure as given below:
\begin{eqnarray}
\langle N_{s^{\prime}}(p^{\prime})| 
\overline{\psi}(0)\gamma^5 \gamma^\mu \lambda^\chi\psi(0)
|N_{s}(p)\rangle  &=& {u}_{s^{\prime}}(p^{\prime})
\left[G_A^\chi(Q^{2}) \gamma^\mu \; +\; G_P^\chi(Q^{2})\frac{q^\mu}{2M}
  \right. \cr &&
\;\;\;\;\;\;\;\;\;\;\;\;\;\;\;  + \left. G_T^\chi(Q^2) \frac{n^\mu}{2M}\right]\gamma^5u_{s}(p)\,. 
\label{eq:general_decompositionA}  
\end{eqnarray}
Being similar to the evaluation of the $H_T^\chi(Q^2)$ form factors
discussed in the previous Section, the axial-vector form factors
$G_A^\chi(Q^2)$ are obtained in the same framework with the
corresponding expressions as: 
\begin{equation}
G_{A}^{\chi}(Q^{2}) \;=\; \frac{M}{E}\,\,\int dr\,
r^{2}\,\left[j_{0}(|\bm Q|r) \, \mathcal{G}_{0}^\chi(r) 
\;-\frac{1}{\sqrt{2}}  j_{2}(|{\bm Q}|r)\,  \mathcal{G}_{2}^\chi(r)  \right].
\label{eq:GAModel}
\end{equation}
The nucleon matrix element $\mathcal{G}_{0}^{\chi}(r)$ is given in the
SU(3) case as:
\begin{eqnarray}
\mathcal{G}_{0}^{\chi=3,8} (r) & = &
-\sqrt{\frac{1}{3}} \langle D_{\chi3}^{(8)}\rangle_N \mathcal{A}_{0}(r)
\;+\; \frac{1}{3\sqrt{3}}\frac{1}{I_{1}} \langle D_{\chi8}^{(8)}
J_{3}\rangle_N \mathcal{B}_{0}(r) \cr
&& \;-\; \sqrt{\frac{1}{3}}\frac{1}{I_{2}} \langle d_{ab3}D_{\chi
  a}^{(8)}J_{b} \rangle_N \mathcal{C}_{0}(r) 
\;- \; \frac{1}{3\sqrt{2}}\frac{1}{I_{1}} \langle D_{\chi3}^{(8)} \rangle_N
\mathcal{D}_{0}(r) \cr
&& \;-\; \frac{2}{3\sqrt{3}} \frac{K_{1}}{I_{1}}M_{8} \langle D_{83}^{(8)}
 D_{\chi8}^{(8)} \rangle_N 
 \mathcal{B}_{0}(r) + \frac{2}{\sqrt{3}}\frac{K_{2}}{I_{2}}M_{8} \langle
 d_{ab3}  D_{8a}^{(8)}D_{\chi b}^{(8)} \rangle_N \mathcal{C}_{0}(r) \cr
 &  & -\sqrt{\frac{1}{3}}\left[2M_{1} \langle D_{\chi3}^{(8)}
   \rangle_N +
   \frac{2}{\sqrt{3}}M_{8} \langle D_{88}^{(8)} D_{\chi3}^{(8)}
   \rangle_N \right] \mathcal{H}_{0}(r) \cr
 &  & +\frac{2}{3\sqrt{3}}M_{8} \langle D_{83}^{(8)}D_{\chi8}^{(8)}
 \rangle_N  \mathcal{I}_{0}(r) - \frac{2}{\sqrt{3}}M_{8} \langle
 d_{ab3}D_{\chi a}^{(8)} D_{8b}^{(8)} \rangle_N \mathcal{J}_{0}(r)\, ,
\label{eq:3,8CQSMdensityA}\\
\mathcal{G}_{0}^0 (r) &=& \frac{1}{3}\frac{1}{I_{1}} \langle
J_{3}\rangle_N \mathcal{B}_{0}(r) -  \frac{2}{3}\frac{K_{1}}{I_{1}}
M_{8} \langle D_{83}^{(8)}\rangle_N  \mathcal{B}_{0}(r) +
\frac{2}{3}M_{8} \langle D_{83}^{(8)} \rangle_N  \mathcal{I}_{0}(r)\,,   
\label{eq:0CQSMdensityA} 
\end{eqnarray}
where the densities $\mathcal{A}_{0},\cdots, \mathcal{J}_{0}$ are related to those
of the tensor form factor by simply dropping the $\gamma_4$ appearing in the
densities $\mathcal{A}_{T0},\cdots, \mathcal{J}_{T0}$ given in the
Appendix. 

The axial-vector form factors in the SU(2) $\chi$QSM is given as
\begin{eqnarray}
\mathcal{G}_{0}^{3} (r) & = &
-\sqrt{\frac{1}{3}} \langle D_{33}\rangle_N \mathcal{A}_{0}(r)
\;- \; \frac{1}{3\sqrt{2}}\frac{1}{I_{1}} \langle D_{33} \rangle_N
\mathcal{D}_{T0}(r) \label{eq:3CQSMdensitySU2A} \\
\mathcal{G}_{0}^0(r) &=& \frac{1}{3}\frac{1}{I_{1}} \langle
J_{3}\rangle_N \mathcal{B}_{0}(r).
\label{eq:0CQSMdensitySU2A} 
\end{eqnarray}
The above given expressions for the axial-vector form factors are
equivalent to those obtained in \cite{Silva:AxialFF}. However, the
given numerical results in the present work are obtained by taking
$\lambda^0=1$ whereas those of the work \cite{Silva:AxialFF}
correspond to taking $\lambda^0=\sqrt{2/3}$. 

%
%
%
%

\section{Tensor and Axial-Vector Charges}
In the case of the tensor and axial-vector charges, i.e. the form
factors at $Q^2=0$, it is possible to write the corresponding
expressions in a very compact way. At the point $Q^2=0$ the second
terms in Eqs. (\ref{eq:HTModel},\ref{eq:GAModel}) vanish and the
spherical Bessel function of the first terms is reduced to unity.  
Hence, all the model-dependent dynamical parts are just given by
the integerals such as $\int dr r^2 \mathcal{A}$, which are just
simple numbers. The residual factors such as $\langle D_{33}^{(8)}
\rangle_N$ are SU(3) Clebsch-Gordan coefficients which can be
derived by the expression given in the Appendix.

The expressions Eqs. (\ref{eq:HTModel},\ref{eq:GAModel}) for the
tensor and axial-vector charges can 
be reduced to the following expressions in SU(3):
\begin{eqnarray}
g^{\chi=3,8} & = &
-\frac{ \langle D_{\chi3}^{(8)}\rangle_N }{\sqrt{3}} \; A
\;+\; \frac{\langle D_{\chi8}^{(8)}J_{3}\rangle_N}{3I_{1}\sqrt{3}}\; B \;-\;
\frac{\langle d_{ab3}D_{\chi a}^{(8)}J_{b} \rangle_N}{I_{2}\sqrt{3}}\; C 
\;- \; \frac{\langle D_{\chi3}^{(8)} \rangle_N}{3I_{1}\sqrt{2}}\;  D \cr
&& \;-\;M_{8} \frac{2K_{1}\langle D_{83}^{(8)} D_{\chi8}^{(8)}
  \rangle_N}{3I_{1}\sqrt{3}} B +M_{8} \frac{2K_{2}\langle 
 d_{ab3}  D_{8a}^{(8)}D_{\chi b}^{(8)} \rangle_N}{3I_{2}}\; C \cr
 &  & -\sqrt{\frac{1}{3}}\left[2M_{1} \langle D_{\chi3}^{(8)}
   \rangle_N +
   \frac{2}{\sqrt{3}}M_{8} \langle D_{88}^{(8)} D_{\chi3}^{(8)}
   \rangle_N \right]\; H \cr
 &  & +M_{8}\frac{2\langle D_{83}^{(8)}D_{\chi8}^{(8)}
   \rangle_N}{3\sqrt{3}}\; I -M_{8} \frac{2 \langle  d_{ab3}D_{\chi
     a}^{(8)} D_{8b}^{(8)} \rangle_N }{\sqrt{3}} \;  J \, , 
\label{eq:3,8CQSMchargeSU3}\\
g^0 &=& \frac{\langle J_{3}\rangle_N}{3I_{1}}\; B - M_{8} \frac{2K_{1}\langle
  D_{83}^{(8)}\rangle_N }{3I_{1}} \; B + M_8 \frac{2\langle D_{83}^{(8)}
  \rangle_N}{3} \;  I\,,   
\label{eq:0CQSMchargeSU3} 
\end{eqnarray}
and those in SU(2):
\begin{eqnarray}
g^{3} & = & -\frac{\langle D_{33}\rangle_N}{\sqrt{3}}  A\;- \;
\frac{\langle D_{33} \rangle_N}{3I_{1}\sqrt{2}}  D \label{eq:3CQSMchargeSU2}  \\
g^0 &=& \frac{\langle J_{3}\rangle_N}{3I_{1}} B.
\label{eq:0CQSMchargeSU2} 
\end{eqnarray}

We list in Table~\ref{tab:denINTEGRATED} in the Appendix   
the values for the densities $\mathcal{A}_{T0},\cdots,
\mathcal{J}_{T0}$ and $\mathcal{A}_{0},\cdots, \mathcal{J}_{0}$
integrated over $r$ with the weight $r^2$as obtained in the
$\chi$QSM. In the case of SU(3) symmetry, i.e. without $m_s$
corrections, the nucleon matrix elements as used in 
Eqs. (\ref{eq:3,8CQSMchargeSU3},\ref{eq:0CQSMchargeSU3}, 
\ref{eq:3CQSMchargeSU2},\ref{eq:0CQSMchargeSU2})
are given in Table~\ref{tab:nucleonMATRIXelements} in the
Appendix. All results at $Q^2=0$ as given in the present work can be
reproduced by using the results listed in
Tables~\ref{tab:denINTEGRATED},\ref{tab:nucleonMATRIXelements}
together with Eqs. (\ref{eq:3,8CQSMchargeSU3},\ref{eq:0CQSMchargeSU3},
\ref{eq:3CQSMchargeSU2},\ref{eq:0CQSMchargeSU2}). 

%
%
%
%

\section{Results and Discussion}
We are now in a position to discuss the results for the tensor form
factors $H_{T}(Q^{2})$ of Eq.~(\ref{eq:general_decomposition}). 
We want to mention that all model parameters are the same as in   
the former works~\cite{AxialTheta,CQSM:ADEFF,CQSM:SHD, 
Silva:AxialFF,Silva:G0A4Happex,Silva:ParityViol,Silva:StrangeFF,
CQSM:Theta_VFF}. For a given $M$, the regularization cut-off parameter
$\Lambda$ and the current quark mass $\overline{m}$ in the Lagrangian
are then fixed to the pion decay constant $f_{\pi}$ and the pion mass
$m_{\pi}$, respectively. Throughout this work the strange current
quark mass is fixed to $m_{\mathrm{s}}=180\,\textrm{MeV}$ which
approximately reproduces the kaon mass. Hence, we do not have any
further free parameter for the present investigation. Especially, 
this is the merit of the $\chi$QSM which enables us to investigate all
baryon form factors within exactly the same framework. In the present
case, these are the tensor and axial-vector form factors. We also
apply the symmetry-conserving quantization as found in
Ref.~\cite{SymmetryConQuantization}. The experimental proton 
electric charge radius is best reproduced in the $\chi$QSM with the
constituent quark mass $M=420$ MeV which is thus our preferred
value. Nevertheless we have checked in this work that the results for
the tensor form factors are rather stable with $M$ varied, so that we
present all results with $M=420$ MeV. 

\begin{table}[h]
\caption{\label{tab:tensor_charge} Tensor charges in comparison with
the axial-vector charges. Both charges, in SU(2) and SU(3), have
been calculated with the same set of parameters in the present work.
The results for the non-relativistic quark model are also given.}    
\begin{center}
\begin{tabular}{c|ccc|ccc||cc|cc|cc}\hline\hline
& $g_{T}^{0}$ &  $g_{T}^{3}$ & $g_{T}^{8}$ & $g_A^{0} $ & $g_A^3$ &
$g_A^8$&$ \Delta u$&$ \delta u$&$ \Delta d$&$\delta d$&$\Delta
s$&$\delta s $ \tabularnewline  
\hline
$\chi$QSM SU(3) & $0.76$ & $1.40$ & $0.45$ & $0.45$  & $1.18$ &
$0.35$&$0.84$&$1.08$&$-0.34$&$-0.32$&$-0.05$&$-0.01$ 
\tabularnewline 
$\chi$QSM SU(2) & $0.75$ & $1.44$ & $--$ & $0.45$  & $1.21$ & $--$&
$0.82$&$1.08$&$-0.37$&$-0.32$&$--$&$--$  
\tabularnewline 
NRQM & $1$ & $5/3$ & $--$ & $1$  & $5/3$ & $--$
&$\frac{4}{3}$&$\frac{4}{3}$&$-\frac{1}{3}$&$-\frac{1}{3}$&$--$&$--$ 
\tabularnewline 
\hline\hline
\end{tabular}
\end{center}
\end{table}

We first concentrate on the tensor charges $g_T^\chi = H_{T}^\chi(0)$
for the singlet and non-singlet components corresponding respectively
to $\chi=0$ and $\chi=3,8$. In Table~\ref{tab:tensor_charge} we list
the results of the tensor charges for the singlet and nonsinglet
components, comparing them with those of the axial-vector
charges. In Ref.~\cite{TensorSU2} the SU(2) isovector and isosinglet
tensor charges were obtained to be $g_T^3=1.45$ and
$g_T^0=0.69$ which are in agreement with those obtained in the present
work.  From the above results, we find the following interesting
inequalities  
\begin{equation}
g_T^\chi > g_A^\chi,
\label{eq:inequal}
\end{equation}
that is, the tensor charges turn out to be in general larger than the
axial-vector charges. These inequalities are also true for the SU(2) 
results.  As mentioned already, the only difference between the tensor
and axial-vector operators is due to the factor of
$\gamma_4$. Therefore, both charges coincide in the  
non-relativistic limit~\cite{Jaffe:tensor,HeJi:tensor}. As discussed
already in Ref.~\cite{TensorSU2}, this can be qualitatively understood
from the asymptotics of both charges in soliton size $R_0$.  The tensor 
charges show generally weaker dependence on the soliton size than the
axial-vector ones do~\cite{TensorSU2}:
\begin{eqnarray}
g_A^3 \sim (MR_0)^2 &\;\;\;\;& g_T^3 \sim MR_0 \cr
g_A^0 \sim \frac{1}{(MR_0)^4} &\;\;\;\;& g_T^0 \sim
\frac{1}{MR_0}\,\,\,.  
\end{eqnarray}
As a result, the tensor charges from the $\chi$QSM turn out to be
closer to those from the nonrelativistic quark model (NRQM) (the limit
of the small soliton size $R_0\to 0$) than the corresponding
axial-vector charges. A similar conclusion was drawn in the bag model
\cite{Jaffe:tensor}.  

The tensor charges were studied independently within the
SU(2) $\chi$QSM in Ref.~\cite{Wakamatsu1999:H(x00),Wakamatsu3:H(x00)}
in which the nucleon structure functions have been calculated. The
tensor and axial-vector charges were derived as the first moments of
the longitudinally and transversely polarized distribution functions,
respectively. Reference~\cite{Wakamatsu1999:H(x00)} obtained the SU(2)
axial-vector charges as $g_A^0=0.35$ and $g_A^3=1.41$, and the tensor
charges as $g_T^0=0.56$ and $g_T^3=1.22$ whereas
Ref.~\cite{Wakamatsu3:H(x00)} used $g_A^0=0.35$,
$g_A^3=1.31$, $g_T^0=0.68$ and $g_T^3=1.21$. While the values for
the singlet from Ref.~\cite{Wakamatsu1999:H(x00),Wakamatsu3:H(x00)}
are similar to those of the present work, the values for $g_A^3$ and
$g_T^3$ seem to be somewhat different. Moreover, their results of
$g_A^3$ and $g_T^3$ do not show the inequality of
Eq.~(\ref{eq:inequal}). However, note that their ratios of 
$g_A^0/g_A^3\simeq 0.25(0.27)$ and $g_T^0/g_T^3\simeq 0.46 (0.56)$ and
the SU(2) ratios of the present work $g_A^0/g_A^3= 0.37$ and
$g_T^0/g_T^3= 0.52$ show the same deviation from the non-relativiestic
quark model value $g^0/g^3=0.6$. 

In the case of the SU(3) $\chi$QSM, the tensor charges were already
studied in \cite{TensorSU3}. However, the former calculation was done without
the symmetry-conserving quantization which ensures the correct realization of
the Gell-Mann-Nishijima formula, and yielded the
following results: $g_T^0=0.70$, $g_T^3=1.54$, and
$g_T^8=0.42$. Compared to the present results listed in
Table~\ref{tab:tensor_charge}, the previous results are deviated from    
the present ones by about $6\sim 10\,\%$.       

The singlet and nonsinglet tensor charges can be decomposed 
into the tensor charges for each flavor as follows: 
\begin{eqnarray} 
\delta u & = & \frac{1}{2}\left(\frac{2}{3} g_{T}^{0} + g_{T}^{3}
  + \frac{1}{\sqrt{3}} g_{T}^{8}\right)\,,\cr
\delta d & = & \frac{1}{2}\left(\frac{2}{3} g_{T}^{0} - g_{T}^{3}
  + \frac{1}{\sqrt{3}} g_{T}^{8} \right)\,,\cr
\delta s & = & \frac{1}{3}\left(g_{T}^{0} - \sqrt{3} g_{T}^{8}\right)\,.
\label{eq:falvor_tensor}
\end{eqnarray}

In Table~\ref{tab:tensor_charge}, we list the results for the
flavor-decomposed tensor charges of the nucleon in comparison with the
corresponding axial-vector ones. As in the case of 
the axial-vector charges, the rotational $1/N_c$ corrections are also
crucial for the tensor charges. On the other hand, the $m_{s}$
corrections that come from both the operators and wave function
corrections turn out to be rather small, i.e. below $5\,\%$.
Moreover, the Dirac-sea quark contribution to the form factor
$H_{T}(Q^{2})$ is almost negligible.  The strange tensor charge turns
out to be tiny. Compared to the work~\cite{Wakamatsu3:H(x00)}, 
Table~\ref{tab:tensor_charge} shows the same tendency, namely: $\delta
u>\Delta u$ and $\delta d > \Delta d$. 

\begin{table}[ht]
\begin{center}
\caption{\label{tab:dd/du}The scale independent quantity $|\delta
d/\delta u|$. References \cite{TensorSU2}, \cite{Wakamatsu3:H(x00)}, 
and \cite{Lorce:tensor} correspond respectively to the the SU(2)
$\chi$QSM, the same model by Wakamatsu, and the SU(3) infinite
momentum frame $\chi$QSM. The values of the works
\cite{HeJi:tensor,Pasquini:2005,Schimdt:HOmodel,Pasquini:CQMschlumpf}  
were obtained in the SU(6) symmetric CQM. The SU(6) symmetric Ansatz 
induces a ratio of $1/4$. Reference~\cite{Anselmino:2009} 
represents a global analysis of SIDIS experimental data. }
\begin{tabular}{c|cccccccc}\hline\hline
Proton & This work &
SU(2)\cite{TensorSU2} & Ref.~\cite{Wakamatsu3:H(x00)} &
IMF\cite{Lorce:tensor} &
CQM\cite{HeJi:tensor,Pasquini:2005,Schimdt:HOmodel,Pasquini:CQMschlumpf} 
& Lattice\cite{Goeckeler:GPDlattice} & SIDIS\cite{Anselmino:2009} &
NR\tabularnewline 
\hline
$|\delta d/\delta u|$ & $0.30$ & $0.36$ & $0.28$ & $0.27$ & $0.25$ &
$0.25$ & $0.42_{-0.20}^{+0.0003}$ & $0.25$\tabularnewline\hline\hline 
\end{tabular}
\end{center}
\end{table}
Table~\ref{tab:dd/du} lists the ratio of the tensor charges $|\delta
d/\delta u|$ for the proton, compared to different approaches.  Note
that this ratio is independent of any renormalization
scale~\cite{Wakamatsu:tensorGPD}. For comparision, we first consider
the results of the SU(2) $\chi$QSM~\cite{TensorSU2} as well as those
of the same model by Ref.~\cite{Wakamatsu3:H(x00)}.  We also compare
the present results with those of the SU(3) infinite-momentum frame
$\chi$QSM~\cite{Lorce:tensor} and of the 
following SU(6)-symmetric model: MIT-bag model~\cite{HeJi:tensor}, the
harmonic oscillator (HO) and hypercentral (HYP) light-cone quark 
models of \cite{Schimdt:HOmodel,Pasquini:2005}. Furthermore,
Ref.~\cite{Pasquini:CQMschlumpf} followed the approach of
Ref.~\cite{Pasquini:2005} by using a proton wave function derived in 
a quark model~\cite{Schlumpf:CQM}. The results of the
CQM~\cite{Schimdt:HOmodel,Pasquini:2005,Pasquini:CQMschlumpf} 
correspond to the three valence quark contribution without the Dirac
sea in the framework of the light-cone quantization. The
approximations used in the present work and in the SU(3) infinite
momentum framework $\chi$QSM of Ref.~\cite{Lorce:tensor} are quite
opposite each other. In Ref.~\cite{Lorce:tensor} the rotation of the
$\chi$QSM soliton can be taken exactly while the Dirac-sea
contribution is truncated. In the present formalism the whole
Dirac-sea contribution 
is included while the rotation of the soliton is taken
perturbatively. In the case of the SU(6)-symmetric models
\cite{HeJi:tensor,Pasquini:2005,Schimdt:HOmodel,Pasquini:CQMschlumpf},
the SU(6)-symmetric Ansatz induces a ratio $|\delta d/\delta 
u|=1/4$.  As for the experimental value of $|\delta d/\delta u|$ we
take the results of Ref.~\cite{Anselmino:2009}, where $\delta
u=0.54_{-0.22}^{+0.09}$ and $\delta d=-0.23_{-0.16}^{+0.09}$ at
$\mu^{2}=0.8\,\mathrm{GeV}^{2}$ were obtained. Compared to 
that value, all theoretical results look underestimated, however,
still within uncertainty. 

From the results of the SU(2) and SU(3) $\chi$QSM listed in
Table~\ref{tab:dd/du}, we find that the SU(2) result from
Ref.~\cite{TensorSU2} seems deviated from that of
Ref.~\cite{Wakamatsu3:H(x00)} and also from that of the present
work. The difference of the present work to Ref.~\cite{TensorSU2} is
mainly due to the fact that the $g_T^0$ differ by a value of $0.06$,
i.e. $8$\%. This difference could be explained by the fact that the
soliton profile and discretization parameters used in
Ref.~\cite{TensorSU2} are different from the present work. Taking into
account this, a deviation of $8$\% is an acceptable one.  
In comparisson to the work Ref.~\cite{Wakamatsu3:H(x00)} the ratio $\delta d
/\delta u$ is comparable to the present one since both the $\delta d$ and
$\delta u$ are approximately by the same factor smaller as compared to the
results of the present work. 

In Ref.~\cite{Pasquini:2005}, the following relation was presented:
\begin{equation}
2\delta u\;=\; \Delta u+\frac{4}{3},\,\,\,\,\, 2\delta
d\;=\;\Delta d-\frac{1}{3}\,,
\end{equation}
which is compatible with the Soffer inequality~\cite{Soffer:1995}.
It is worthwhile to note that this relation is numerically
approximately fulfilled by the results of the present work.   

In the following, we will compare our results to the lattice
calculation Ref.~\cite{Goeckeler:GPDlattice}. The lattice results for
the tensor form factors were derived at a renormalization scale of 
$\mu^{2}=4\,\mathrm{GeV}^{2}$ and are linearly extrapolated to the
physical pion mass as well as to the continuum. Disconnected
quark-loop diagrams were not considered. In the $\chi$QSM, the
renormalization scale is given by the cut-off mass of 
the regularization, which is approximately $\mu^{2}\approx
0.36\,\mathrm{GeV}^{2}$.  This value is related implicitly to the  
size of the QCD instantons 
$\overline{\rho}\approx0.35\,\mathrm{fm}$~\cite{Diakonov:1983hh,
  Diakonov:1985eg}.
We use Eq.~(\ref{eq:evolve}) in order to evolve the lattice and SIDIS
results~\cite{Anselmino:2009} from $4\,\mathrm{GeV}^{2}$ and
$0.80\,\mathrm{GeV}^{2}$, respectively, to the renormalization scale
of the $\chi$QSM $0.36\,\mathrm{GeV}^{2}$. We compare the present
results with those of the lattice calculations as follows: 
\begin{eqnarray}
\mbox{SIDIS \cite{Anselmino:2009} ($0.80\,\mathrm{GeV}^{2}$):} &  &
\delta 
u=0.54^{+0.09}_{-0.22}\,,\,\,\,\,\,\,\,\,\,\,\,\,\,\,\,\,\,\,\,\,\,\,\,\,\delta  
d=-0.231^{+0.09}_{-0.16},\cr
\mbox{SIDIS \cite{Anselmino:2009} ($0.36\,\mathrm{GeV}^{2}$):} &  &
\delta 
u=0.60^{+0.10}_{-0.24}\,,\,\,\,\,\,\,\,\,\,\,\,\,
\,\,\,\,\,\,\,\,\,\,\,\,\delta 
d=-0.26^{+0.1}_{-0.18},\cr
\mbox{Lattice \cite{Goeckeler:GPDlattice} ($4.00\,\mathrm{GeV}^{2}$):}
&  & \delta 
u=0.86\pm0.13\,,\,\,\,\,\,\,\,\,\,\,\,\,\,\,\,\,\,\,\delta
d=-0.21\pm0.005\,,\cr
\mbox{Lattice \cite{Goeckeler:GPDlattice} ($0.36\,\mathrm{GeV}^{2}$):}
&  & \delta 
u=1.05\pm0.16\,,\,\,\,\,\,\,\,\,\,\,\,\,\,\,\,\,\,\,\delta 
d=-0.26\pm0.01\,,\cr
\chi\mbox{QSM ($0.36\,\mathrm{GeV}^{2}$):} &  & \delta
u=1.08\,,\,\,\,\,\,\,\,\,\,\,\,\,\,\,\,\,\,\,\,\,\,\,\,
\,\,\,\,\,\,\,\,\,\,\,\,\,\delta d=-0.32\,, \nonumber
\end{eqnarray} 
from which we find that the results are generally in a good agreement
with those of the lattice
calculation~\cite{Goeckeler:GPDlattice}. However, both approaches  
disagree with the SIDIS results~\cite{Anselmino:2009} for $\delta 
u$ by nearly a factor of 2. 

\begin{figure}[ht]

\begin{center}

\includegraphics[scale=0.60]{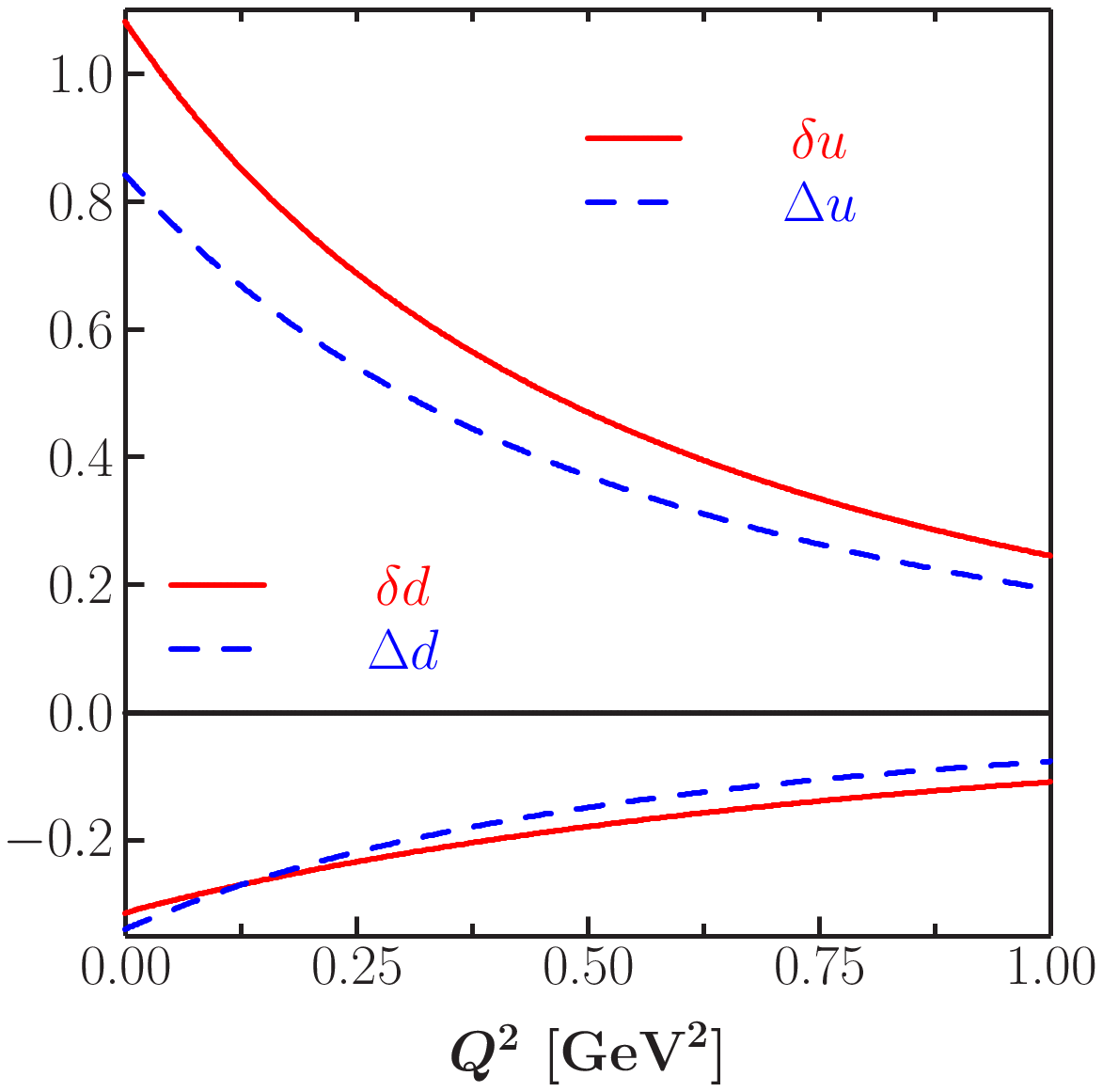}
\includegraphics[scale=0.60]{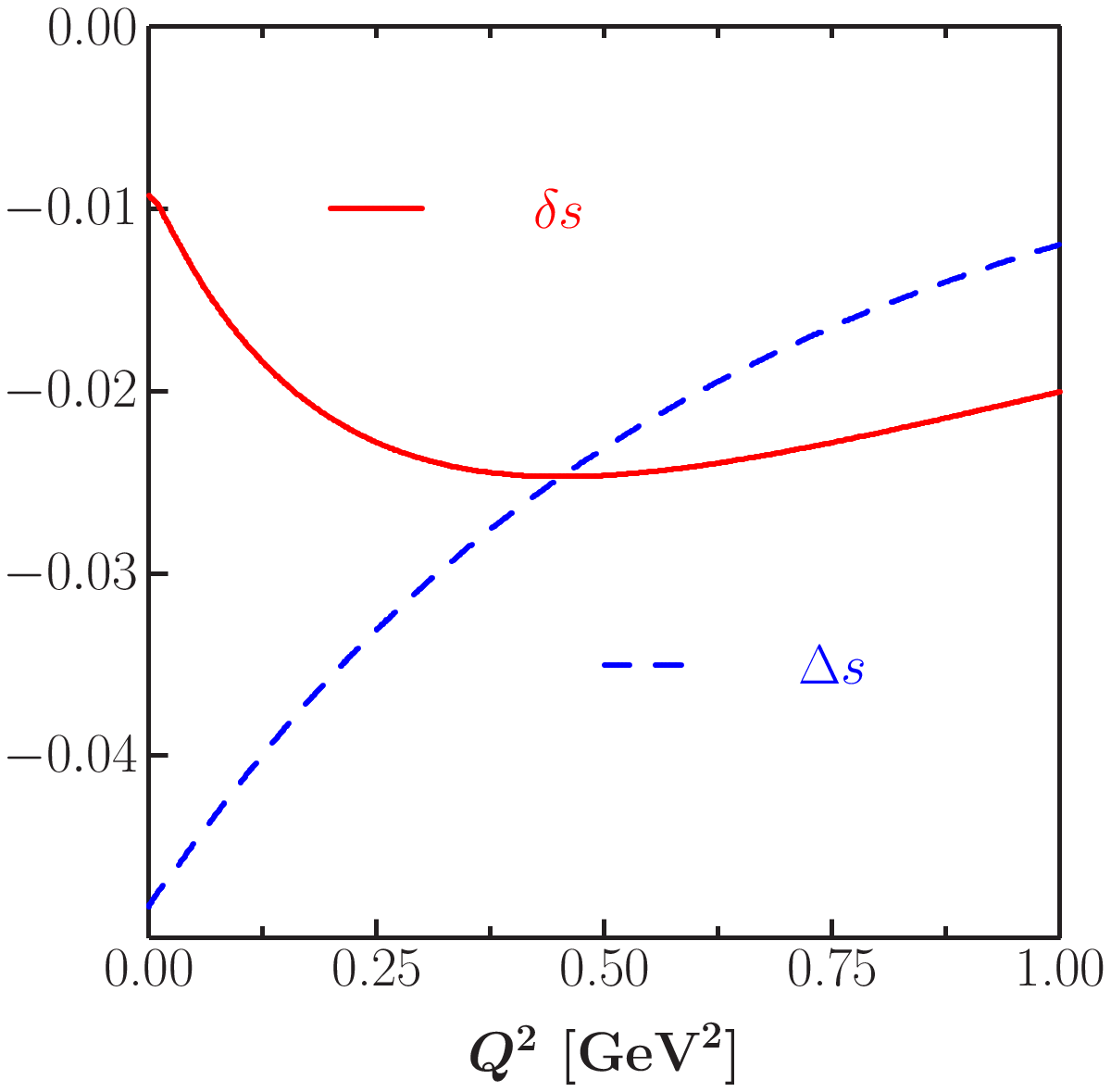}\\
\includegraphics[scale=0.60]{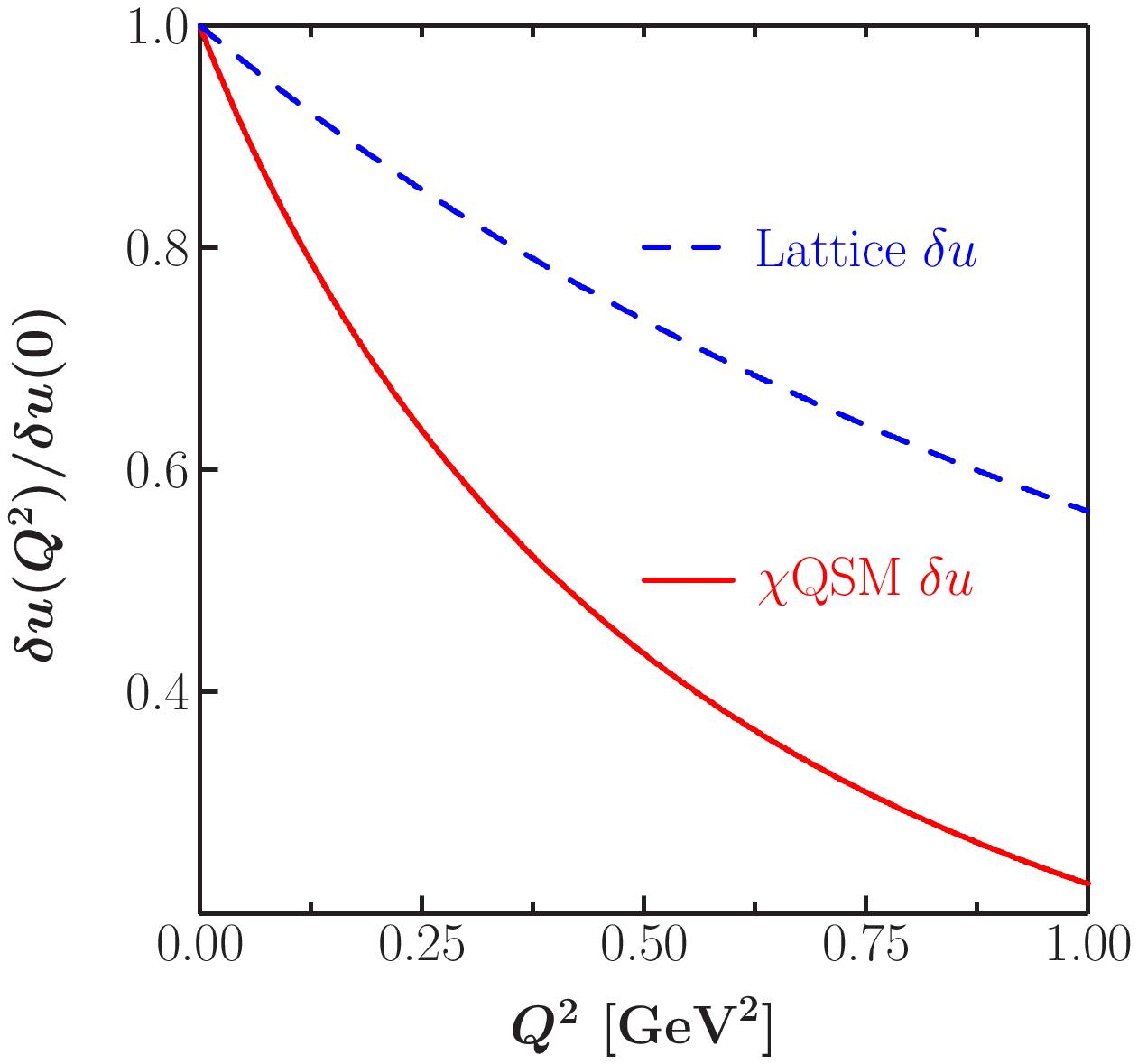}
\includegraphics[scale=0.60]{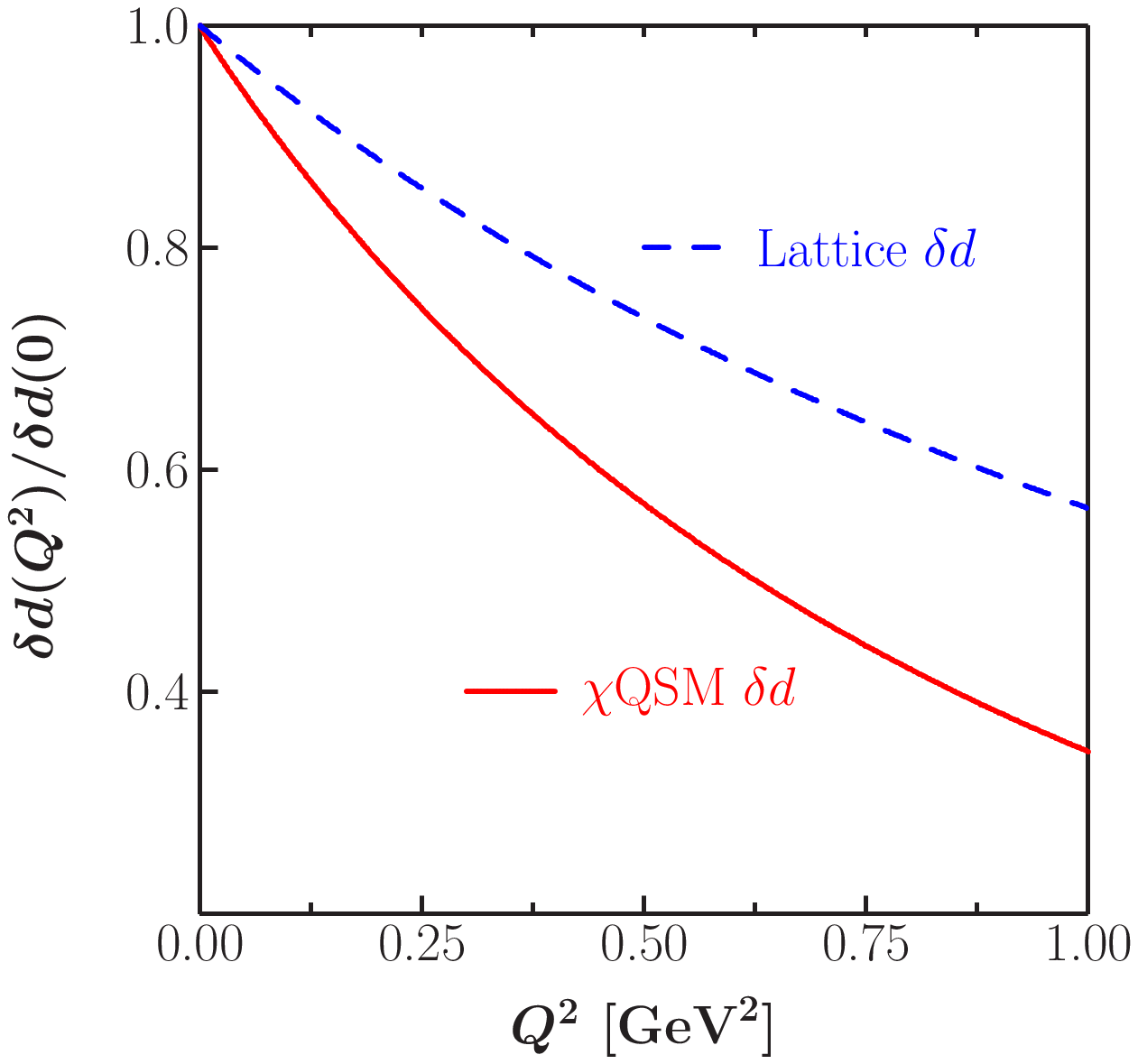}

\end{center}
\caption{\label{Fig: HT}(Color online) flavor tensor form factors  
$\delta u(Q^{2})$, $\delta d(Q^{2})$ and $\delta s(Q^{2})$ for the
proton. In the upper panel, we compare the tensor form factors with
the axial-vector ones. The red (solid) curves show the tensor form
factors whereas the blue (dashed) ones represent the axial-vector form 
factors. In the lower panel, we compare the present results of the
renormalization-independent scaled tensor form factors $\delta
u(Q^{2})/\delta u(0)$, $\delta d(Q^{2})/\delta d(0)$ with those of the
lattice QCD~\cite{Goeckeler:GPDlattice}. The red (solid) curves
designate the $\chi$QSM form factors of this work while the blue
(dashed) ones corrrespond to the factors from the lattice QCD
calculation.}    
\end{figure}

We  now turn to the tensor form factors $H_{T}^\chi(Q^{2})$ calculated
up to the momentum transfer of $Q^{2}\leq1\,\mathrm{GeV}^{2}$. The
tensor form factors $H_{T}^\chi(Q^{2})$ expressed in
Eq.~(\ref{eq:HTModel}) consists of two densities. While only the first
one determines the tensor charges at $Q^{2}=0$, both densities are
responsible for $H_{T}^\chi(Q^{2})$.  In the upper left panel of
Fig.~\ref{Fig: HT}, the up and down tensor form factors are shown
together with the corresponding axial-vector form factors.
Interestingly, we find that the general behaviors of these form
factors are very similar.  In particular, 
$\delta d$ and $\Delta d$ look pretty much similar to each other.  
In the upper right panel, the strange tensor and axial-vector
form factors are drawn. In contrast to the nonstrange form factors,
the strange tensor form factor seems to be very different from the
axial-vector one. The $Q^2$ dependence of the strange tensor form
factor is somewhat peculiar. It behaves like a neutral form factor.  
Moreover, the strange tensor charge turns out to be very small,
compared to the axial-vector one. It implies that the tensor form
factors are ``less relativistic''.    

In the lattice work~\cite{Goeckeler:GPDlattice} the tensor form
factors $\delta u(Q^{2})$ and $\delta d(Q^{2})$ were calculated up to
a momentum transfer of $Q^{2}=3.5\,\mathrm{GeV}^{2}$. In the lower
panel of Fig. \ref{Fig: HT},  we compare our scaled form factors
$\delta q(Q^2)/\delta q(0)$ with those of the lattice 
calculation~\cite{Goeckeler:GPDlattice}. In fact, these scaled form
factors are independent of the renormalization scale.  
While the lattice result for $\delta u$ decreases almost linearly as
$Q^2$ increases, that of the present work falls off more 
rapidly. Thus, the value of $\delta u$ at $Q^2=1\,\mathrm{GeV}^2$ is
almost $2.5$ times smaller than that of the lattice
calculation.  Note, however, that this is an expected result.  A
similar behavior was also obtained for the $\Delta(1232)$ electric
quadrupole form factor as shown in Ref.~\cite{CQSM:DeltaEQM}. These
differences of the $Q^2$ dependence can be understood from the fact
that lattice calculations tend to yield rather flat form factors
because of the heavy pion mass employed. This has been shown
explicitly for the nucleon isovector form factor $F_{1}^{p-n}(Q^{2})$
on the lattice~\cite{Lattice:FuMd}.  

In general, the form factors from the $\chi$QSM are well 
reproduced by the dipole formula
\begin{eqnarray}
H_{T}(Q^{2}) & = &
\frac{H_{T}(0)}{(1+Q^{2}/ M_{\mathrm{d}}^2)^{2}}
\label{EQ:HTdipoleFIT}
\end{eqnarray}
with  the dipole mass $M_{\mathrm{d}}$. A direct fit leads to the 
dipole masses corresponding to the tensor form factors for
$\chi=0,3,8$ and the up and down form factors $\delta u(Q^{2})$ and
$\delta d(Q^{2})$ as listed in Table~\ref{Tab. DipoleM}. Note that,
however, the strange tensor form factors cannot be fitted in terms of
the dipole type.  
\begin{table}[h]
\caption{\label{Tab. DipoleM} Dipole masses $M_{\mathrm{d}}$ for
  the tensor form factors $H_{T}^{\chi}(Q^{2})$, $\delta u(Q^{2})$ and
  $\delta d(Q^{2})$. Given are the values of the tensor form
  factors for each flavor at $Q^{2}=0$ and the dipole 
  masses in units of $\mathrm{GeV}$, which reproduce the present
  results.} 
\begin{center}
\begin{tabular}{c|ccccccc}\hline\hline
Proton & $H_{T}^{0}(Q^{2})$ & $H_{T}^{3}(Q^{2})$ & $H_{T}^{8}(Q^{2})$
&  &  & $\delta u(Q^{2})$ & $\delta d(Q^{2})$\tabularnewline 
\hline
$Q^{2}=0$ & $0.76$ & $1.40$ & $0.45$ &  &  & $1.08$ &
$-0.32$\tabularnewline 
$M_{\mathrm{d}}$ & $0.851$ & $1.03$ & $0.984$ &  &  & $0.980$
& $1.24$\tabularnewline 
\hline\hline
\end{tabular}

\end{center}
\end{table}

\begin{table}[h]

\caption{\label{tab:octet} Tensor charges for $\delta q(0)$ for the
  baryon octet.}

\begin{center}

\begin{tabular}{c|rrrrrrrr}\hline\hline
 & $p(uud)$ & $n(ddu)$ & $\Lambda(uds)$ & $\Sigma^{+}(uus)$ &
 $\Sigma^{0}(uds)$ & $\Sigma^{-}(dds)$ & $\Xi^{0}(uss)$ &
 $\Xi^{-}(dss)$\tabularnewline 
\hline
$\delta u$ & $1.08$ & $-0.32$ & $-0.03$ & $1.08$ & $0.53$ & $-0.02$ &
$-0.32$ & $-0.02$\tabularnewline 
$\delta d$ & $-0.32$ & $1.08$ & $-0.03$ & $-0.02$ & $0.53$ & $1.08$ &
$-0.02$ & $-0.32$\tabularnewline 
$\delta s$ & $-0.01$ & $-0.01$ & $0.79$ & $-0.29$ & $-0.29$ & $-0.29$
& $1.06$ & $1.06$\tabularnewline 
 \hline\hline
\end{tabular}

\end{center}
\end{table}

For completeness, we list in Table~\ref{tab:octet} the tensor charges 
$\delta q$ for the baryon octet. Having scrutinized the results in 
Table~\ref{tab:octet}, we find the following relations:  
\begin{eqnarray}
\delta u_p &=& \delta d_n,\;\;\;\;\;\;\,\delta u_n \;=\; \delta
d_p,\;\;\;\;\;\;\, 
\delta u_{\Lambda}\;=\; \delta d_{\Lambda},\;\;\;\;\, \delta u_{\Sigma^+}
\;=\; \delta d_{\Sigma^-},\cr
\delta u_{\Sigma^0} &=& \delta d_{\Sigma^0},\;\;\;
\delta u_{\Sigma^-} \;=\; \delta d_{\Sigma^+},\;\;\; \delta
u_{\Xi^0}\;=\; \delta d_{\Xi^-},\;\;\;\delta u_{\Xi^-}\;=\; \delta
d_{\Xi^0}, \cr
\delta s_{p} &=& \delta s_{n},\;\;\;\;\;\delta s_{\Sigma^\pm} \;=\; \delta
s_{\Sigma^0},\;\;\;\; \delta s_{\Xi^0}\;=\; \delta s_{\Xi^-},
\label{eq:octten}
\end{eqnarray}
which are the consequence of the assumed isospin symmetry in the
present work. Generally, in $SU(3)$ flavor symmetry we have the
following relations:  
\begin{eqnarray}
\delta u_p &=& \delta d_n \;=\; \delta u_{\Sigma^+} \;=\; \delta
d_{\Sigma^-} \;=\; \delta s_{\Xi^0}\;=\; \delta s_{\Xi^-},\cr
\delta u_n &=& \delta d_p \;=\; \delta u_{\Xi^0}\;=\; \delta d_{\Xi^-}
\;=\; \delta s_{\Sigma^\pm} \;=\; \delta s_{\Sigma^0}. 
\label{eq:su3ten}
\end{eqnarray}
By comparing the above relations with the numbers given in
Table~\ref{tab:octet} we can see the overall smallness of SU(3) symmetry
breaking contributions for the tensor charges of all octet baryons.
Similar relations to Eqs. (\ref{eq:octten}, \ref{eq:su3ten}) can be
found also for the axial-vector charges and magnetic moments of the
octet baryon~\cite{axial,mag}. 
%
%
%
%

\section{Summary and conclusion}
In the present work we investigated the tensor form factors 
$H_{T}(Q^{2})$ of the SU(3) baryons, which are deeply related to the
chiral-odd generalized parton distribution $H_{T}(x,\xi,t)$.  We used
the SU(3) self-consistent chiral quark-soliton model ($\chi$QSM) with
symmetry-conserving quantization in order to calculate the tensor
charges and form factors up to the momentum 
transfer $Q^{2}\leq1\,\mathrm{GeV}^{2}$, taking into account linear
rotational $1/N_c$ corrections and linear $m_{\mathrm{s}}$
corrections. All parameters of the model including the constituent
quark mass have been already fixed in reproducing the meson and
nucleon properties. No additional parameter has been fitted in the
present calculation. 

We first computed the flavor singlet and nonsinglet tensor charges of
the nucleon: $g_T^0=0.76$, $g_T^3=1.40$, and $g_T^8=0.45$. 
As for the flavor-decomposed tensor charges $\delta q=H_{T}^{q}(0)$,
we obtained the following results: $\delta u=1.08$, $\delta d=-0.32$
and $\delta s=-0.01$.  We found that for these tensor charges the
Dirac-sea contribution as well as the effects of flavor SU(3) symmetry 
breaking are negligibly small. We compared the present results of the 
tensor form factors $H_{T}^{u,d}(Q^{2})$ with those of the lattice 
QCD~\cite{Goeckeler:GPDlattice}. For the up and down tensor charges, 
i.e. $H_T^{u,d}(0)$, the results are in good agreement with the
lattice data. However, the present results of the tensor form factors
fall off faster than those from the lattice QCD, as $Q^2$ increases.  
The reason for this lies in the fact that the heavier pion mass
utilized in the lattice calculation causes generally flat form
factors. We also presented the tensor charges of the baryon octet. The
results indicated that the effects of SU(3) symmetry breaking turn out
to be negligibly small.   

The second and third tensor form factors, i.e. $E_T(Q^2)$ and
$\tilde{H}_T(Q^2)$, will be discussed elsewhere. The corresponding 
investigation is under way.
\section*{Acknowledgments}
The authors are grateful to P. Hägler, B. Pasquini, P. Schweitzer, and
M. Vanderhaeghen for valuable discussions and critical
comments. T.L. was supported by the Research Centre
``Elementarkr\"afte und Mathematische Grundlagen'' at
the Johannes Gutenberg University Mainz. A.S. acknowledges partial
support from PTDC/FIS/64707/2006. The present work is also supported
by Basic Science Research Program through the National Research
Foundation of Korea (NRF) funded by the Ministry of Education, Science
and Technology (grant number: 2009-0073101). 

%
%
%
%

\begin{appendix}
\section{Densities}
In this Appendix, we provide the densities for the tensor form factors
given in Eq.~(\ref{eq:3,8CQSMdensity}) which
comprise $\mathcal{A}_{T0}(r),...,\mathcal{J}_{T0}(r)$
and $\mathcal{A}_{T2}(r),...,\mathcal{J}_{T2}(r)$. 
The corresponding vector operators $O_1$ in the spherical tensor
operator notation of Ref.~\cite{VMK} for the individual densites are
given as: 
\begin{eqnarray*}
\mbox{for }\mathcal{A}_{T0}(r),...,\mathcal{J}_{T0}(r) & \to &
O_{1}=\sigma_{1}\\ 
\mbox{for }\mathcal{A}_{T2}(r),...,\mathcal{J}_{T2}(r) & \to &
O_{1}=\sqrt{4\pi}\,\,\{Y_{2}\otimes\sigma_{1}\}_{1} 
\end{eqnarray*}
In the following, the sums run freely over all single-quark levels
including valence ones $|v \rangle$ except that the sum over 
$n_0$ is constrained to negative-energy levels:

\begin{eqnarray}
\frac{1}{N_{c}}\mathcal{A}_T(r) & = & \langle
v||r\rangle\gamma_4\{O_{1}\otimes\tau_{1}\}_{0}\langle
r||v\rangle-\frac{1}{2}
\sum_{n}\textrm{sign}(\varepsilon_{n})\sqrt{2G_{n}+1}\langle 
n||r\rangle\gamma_4\{O_{1}\otimes\tau_{1}\}_{0}\langle
r||n\rangle\nonumber \\ 
\frac{1}{N_{c}}\mathcal{B}_T(r) & = &
\sum_{\varepsilon_{n}\neq\varepsilon_{v}}
\frac{1}{\varepsilon_{v}-\varepsilon_{n}}(-)^{G_{n}}\langle 
v||r\rangle\gamma_4 O_{1}\langle r||n\rangle\langle
n||\tau_{1}||v\rangle\nonumber \\ 
 &  & -\frac{1}{2}\sum_{n,m}\mathcal{R}_{3}
 (\varepsilon_{n},\varepsilon_{m})(-)^{G_{m}-G_{n}}\langle
 n||\tau_{1}||m\rangle\langle m||r\rangle\gamma_4 O_{1}\langle
 r||n\rangle\nonumber \\ 
\frac{1}{N_{c}}\mathcal{T}_T(r) & = &
\sum_{\varepsilon_{n^{0}}}\frac{1}{\varepsilon_{v}
  -\varepsilon_{n^{0}}}\langle 
v||r\rangle\gamma_4\{O_{1}\otimes\tau_{1}\}_{0}\langle
r||n^{0}\rangle\langle n^{0}\mid v\rangle\nonumber \\ 
 &  & -\sum_{n,m}\mathcal{R}_{3}(\varepsilon_{n},
 \varepsilon_{m^{0}})\sqrt{2G_{n}+1}\langle
 m^{0}||r\rangle\gamma_4\{O_{1}\otimes\tau_{1}\}_{0}\langle
 r||n\rangle\langle n\mid m^{0}\rangle\nonumber \\ 
\frac{1}{N_{c}}\mathcal{D}_T(r) & = &
\sum_{\varepsilon_{n}}\frac{\textrm{sign}
  (\varepsilon_{n})}{\varepsilon_{v}-\varepsilon_{n}}(-)^{G_{n}}\langle 
v||\tau_{1}||n\rangle\langle
n||r\rangle\gamma_4\{O_{1}\otimes\tau_{1}\}_{1}\langle
r||v\rangle\nonumber \\ 
 &  & +\frac{1}{2}\sum_{n,m}\mathcal{R}_{6}(\varepsilon_{n},
 \varepsilon_{m})(-)^{G_{m}-G_{n}}\langle n||\tau_{1}||m\rangle\langle
 m||r\rangle\gamma_4\{O_{1}\otimes\tau_{1}\}_{1}\langle
 r||n\rangle\nonumber \\ 
\frac{1}{N_{c}}\mathcal{H}_T(r) & = & \sum_{\varepsilon_{n}
  \neq\varepsilon_{v}}\frac{1}{\varepsilon_{v}-\varepsilon_{n}}\langle
v||r\rangle\gamma_4\{O_{1}\otimes\tau_{1}\}_{0}\langle
r|n\rangle\langle n|\gamma^{0}|v\rangle\nonumber \\ 
 &  & -\frac{1}{2}\sum_{n,m}\mathcal{R}_{5}(\varepsilon_{n},
 \varepsilon_{m})\sqrt{2G_{m}+1}\langle
 m||r\rangle\gamma_4\{O_{1}\otimes\tau_{1}\}_{0}\langle
 r||n\rangle\langle n|\gamma^{0}|m\rangle\nonumber \\ 
\frac{1}{N_{c}}\mathcal{I}_T(r) & = & \sum_{\varepsilon_{n}
  \neq\varepsilon_{v}}\frac{1}{\varepsilon_{v} -
  \varepsilon_{n}}(-)^{G_{n}}\langle 
v||r\rangle\gamma_4 O_{1}\langle r||n\rangle\langle
n|\gamma^{0}\tau_{1}||v\rangle\nonumber \\ 
 &  & -\frac{1}{2}\sum_{n,m}\mathcal{R}_{5}
 (\varepsilon_{n},\varepsilon_{m})(-)^{G_{m}-G_{n}}\langle
 n||\gamma^{0}\tau_{1}||m\rangle\langle m||r\rangle\gamma_4
 O_{1}\langle r||n\rangle\nonumber \\ 
\frac{1}{N_{c}}\mathcal{J}_T(r) & = & \sum_{\varepsilon_{n^{0}}}
\frac{1}{\varepsilon_{v}-\varepsilon_{n^{0}}}\langle
v||r\rangle\gamma_4\{O_{1}\otimes\tau_{1}\}_{0}\langle
r||n^{0}\rangle\langle n^{0}|\gamma^{0}|v\rangle\nonumber \\ 
 &  & -\sum_{n,m}\mathcal{R}_{5}(\varepsilon_{n},
 \varepsilon_{m^{0}})\sqrt{2G_{m}+1}\langle
 m^{0}||r\rangle\gamma_4\{O_{1}\otimes\tau_{1}\}_{0}\langle
 r||n\rangle\langle n|\gamma^{0}|m^{0}\rangle\,.
\label{eq:tensor densities A-J thesis}
\end{eqnarray} 
where we take the notation for the reduced matrix elements of the
$r$-dependent states as schematically given below:
\begin{eqnarray*}
\langle n||r\rangle O_{1}\langle r||m\rangle & = &
\left(\begin{array}{cc} 
A(r)\langle i_{n}|| & B(r)\langle
j_{n}||\end{array}\right)O_{1}\left(\begin{array}{c} 
C(r)||i_{m}\rangle\\
D(r)||j_{m}\rangle\end{array}\right)\\
 & = & A(r)C(r)\langle i_{n}||O_{1}||i_{m}\rangle+B(r)D(r)\langle
j_{n}||O_{1}||j_{m}\rangle.
\end{eqnarray*}
The functions $A(r)$, $B(r)$, $C(r)$, $D(r)$ and grand-spin states 
$i_n,i_m,j_n,j_m$ can be found in Ref.~\cite{Wakamatsu:Basis}. 

The regularization functions
$\mathcal{R}_i(\varepsilon_n,\varepsilon_m)$ appearing in
Eq.(\ref{eq:tensor densities A-J thesis}) are given by
\begin{eqnarray*}
\mathcal{R}_{3}(\varepsilon_{n},\varepsilon_{m}) & = &
\frac{1}{2\sqrt{\pi}}\int_{1/\Lambda^{2}}^{\infty}\frac{du}{\sqrt{u}}
\Big[\frac{1}{u} \frac{e^{-\varepsilon_{n}^{2}u} -
  e^{-\varepsilon_{m}^{2}u}}{ \varepsilon_{m}^{2}-\varepsilon_{n}^{2}}
-\frac{\varepsilon_{n}e^{-u\varepsilon_{n}^{2}}+\varepsilon_{m}
  e^{-u\varepsilon_{m}^{2}}}{ \varepsilon_{m}+\varepsilon_{n}} \Big],\\  
\mathcal{R}_{5}(\varepsilon_{n},\varepsilon_{m}) & = & \frac{1}{2}
\frac{\textrm{sign} \varepsilon_{n}-\textrm{sign}
  \varepsilon_{m}}{\varepsilon_{n}-\varepsilon_{m}}, \\ 
\mathcal{R}_{6}(\varepsilon_{n},\varepsilon_{m}) & = &
\frac{1}{2}\frac{1 -\textrm{sign}(\varepsilon_{n})
  \textrm{sign}(\varepsilon_{m})}{ \varepsilon_{n}-\varepsilon_{m}} . 
\end{eqnarray*}

\section{Baryon matrix elements and integrated densities}
All appearing baryon matrix elements in this work are calculated by
using the following relation:
\begin{eqnarray}
 \langle
 B^{\prime}(\mathcal{R}^{\prime})|D_{\chi\alpha}^{(n)}|B(\mathcal{R})\rangle
 &  = &
 \sqrt{\frac{\textrm{dim}(\mathcal{R}^{\prime})}{
     \textrm{dim}(\mathcal{R})}}(-1)^{\frac{1}{2}
   Y_{s}^{\prime}+S_{3}^{\prime}}(-1)^{\frac{1}{2}Y_{s}+S_{3}}  
 \\ 
&&\sum_{\gamma}\left(\begin{array}{ccc}
\mathcal{R}^{\prime} & n & \mathcal{R}_{\gamma}\\
Q_{y}^{\prime} & \chi & Q_{y}\end{array}\right)\left(
\begin{array}{ccc}
\mathcal{R}^{\prime} & n & \mathcal{R}_{\gamma}\\
\overline{Q}_{s^{\prime}} & \alpha &
\overline{Q}_{s}\end{array}\right)\,,
\end{eqnarray}
where $B(\mathcal{R})$ represents a baryon from the SU(3)
representation $R$ with the flavor quantum numbers $Q_{y}=YII_{3}$ and
spin quantum numbers $Q_{s}=Y_{s}SS_{3}$
($\overline{Q_{s}}=-Y_{s}S-S_{3}$). The quantities in brackets
represent the SU(3) Clebsch-Gordan coefficients.
\begin{table}[h]
\caption{\label{tab:nucleonMATRIXelements} SU(3) Nucleon matrix elements}
\begin{center}
\begin{tabular}{lcccccccc}\hline\hline
$\langle D^{(8)}_{33}\rangle_N $&$\langle D^{(8)}_{83}\rangle_N $&$\langle
  D^{(8)}_{38}\rangle_N$&$\langle D^{(8)}_{88}\rangle_N $&$\langle d_{ab3
  }D^{(8)}_{3a}J_b\rangle_N $&$\langle d_{ab3}
  D^{(8)}_{8a}J_b\rangle_N $\\ \hline
$-I_3 \frac{7}{15}$ &$-\frac{\sqrt3}{30}  $& $I_3 \frac{\sqrt3}{15} $ &
  $\frac{3}{10} $ & $I_3 \frac{7}{30}$ &$\frac{\sqrt{3}}{60}$\\
\hline\hline
\end{tabular}
\end{center}
\end{table}

\begin{table}[h]
\caption{\label{tab:denINTEGRATED} Integrated densities for tensor and
  axial-vector charges with the constituent quark mass $M=420$ MeV and
  numerical parameters fixed as described in the text. We use here the
  following notations:  
  $A=\int dr r^2 \mathcal{A}(r)$, $B=\int dr r^2 \mathcal{B}(r)$,
  $C=\int dr r^2 \mathcal{C}(r)$, $D=\int dr r^2 \mathcal{D}(r)$,
  $H=\int dr r^2 \mathcal{H}(r)$, $I=\int dr r^2 \mathcal{I}(r)$,
  $J=\int dr r^2 \mathcal{J}(r)$ }      
\begin{center}
\begin{tabular}{l|cccccccc}\hline\hline
&$A$&$B$&$C$&$D$&$H$&$I$&$J$\\ \hline
Tensor&5.22&4.75&-2.46&5.84&-0.02&2.43&-1.66  \\
Axial-Vector& 4.20&2.86&-2.10&5.46&0.17&1.20&-1.33  \\
\hline\hline
\end{tabular}
\end{center}
\end{table}

\end{appendix}

\end{document}